\documentclass[preprint, longbibliography, 10pt, twocolumn, floatfix, amsmath,amssymb, ]{revtex4-1} 

\usepackage{graphicx}
\usepackage{dcolumn}
\usepackage{bm}
\usepackage{hyperref}

\usepackage{color}

\begin{document}

\title{Persistence Homology of Entangled Rings}

\author{Fabio Landuzzi}
\affiliation{Department of Physics and Mathematics, Aoyama Gakuin University, 5-10-1 Fuchinobe, Chuo-ku, Sagamihara, Japan}

\author{Takenobu Nakamura}
\affiliation{PRESTO, Japan Science and Technology Agency (JST), 4-1-8 Honcho Kawaguchi, Saitama 332-0012, Japan}
\affiliation{National Institute of Advanced Industrial Science and Technology （AIST), 1-1-1, Umezono, Tsukuba, Ibaraki 305-8568, Japan}

\author{Davide Michieletto}
\affiliation{School of Physics and Astronomy, University of Edinburgh, Peter Guthrie Tait Road, Edinburgh, EH9 3FD, United Kingdom}
\affiliation{MRC Human Genetics Unit, Institute of Genetics and Molecular Medicine, University of Edinburgh, North Crewe Rd, Edinburgh, EH4 2XU, United Kingdom}
\affiliation{Centre for Mathematical Biology, and Department of Mathematical Sciences, University of Bath, North Rd, Bath, BA2 7AY, United Kingdom}

\author{Takahiro Sakaue}
\email{sakaue@phys.aoyama.ac.jp}
\affiliation{Department of Physics and Mathematics, Aoyama Gakuin University, 5-10-1 Fuchinobe, Chuo-ku, Sagamihara, Japan}
\affiliation{PRESTO, Japan Science and Technology Agency (JST), 4-1-8 Honcho Kawaguchi, Saitama 332-0012, Japan}

\begin{abstract}
Topological constraints (TCs) between polymers determine the behaviour of complex fluids such as creams, oils and plastics. Most of the polymer solutions used every day life employ linear chains; their behaviour is accurately captured by the reptation and tube theories which connect microscopic TCs to macroscopic viscoelasticity. On the other hand, polymers with non-trivial topology, such as rings, hold great promise for new technology but pose a challenging problem as they do not obey standard theories; additionally, topological invariance -- i.e. the fact that rings must remain unknotted and unlinked if prepared so -- precludes any serious analytical treatment. Here we propose an unambiguous, parameter-free algorithm to characterise TCs in polymeric solutions and show its power in characterising TCs of entnagled rings. We analyse large-scale molecular dynamics (MD) simulations via persistent homology, a key mathematical tool to extract robust topological information from large datasets. This method allows us to identify ring-specific TCs which we call ``homological threadings'' (H-threadings) and to connect them to the polymers' behaviour. It also allows us to identify, in a physically appealing and unambiguous way, scale-dependent loops which have eluded precise quantification so far. We discover that while threaded neighbours slowly grow with the rings' length, the ensuing TCs are extensive also in the asymptotic limit. Our proposed method is not restricted to ring polymers and can find broader applications for the study of TCs in generic polymeric materials.
\end{abstract}


\maketitle

\section{Introduction}

Ring polymers are the simplest class of topologically non-trivial polymers that manifestly depart from the predictions of theoretical cornerstones such as the reptation and tube models~\cite{Doi1988}. Over the last 3 decades, there have been several theoretical and experimental attempts at understanding the statics and dynamics of rings~\cite{Klein1986,McKenna1987,Rubinstein1986,Kapnistos2008,Halverson2011dynamics,Rosa2013,Doi2015,Grosberg2013,Smrek2015a,Bras2014,Ge2016,Sakaue2018,Sakaue2011,Tsalikis2016,Soh2019,OConnor2019,Gomez2020,Sakaue2019,Michieletto2020} and yet their behaviour in entangled solutions is still poorly understood. Individual ring polymers in the melt or entangled solutions assume compact non-Gaussian conformations, which are distinct from the ones assumed by linear polymers. Understanding the topological origin of these conformations and their consequence on the rings entanglements and ensuing dynamics of the bulk is crucial, not only from a fundamental perspective but also for practical applications such as the rheology of ring polymer melts~\cite{Doi2015}, the phase behavior of blends~\cite{Fitzpatrick2018,Chapman2012,Zhou2019,Sakaue2016} and even the dynamical organization of chromosomes in the cell nucleus~\cite{Halverson2013}. 

Due to their unconventional conformations in equilibrium, long non-concatenated rings in the melt experience topological constraints (TCs) that are markedly different from ordinary entanglements between linear chains. Early models for rings polymers pictured them in crumpled, double-folded lattice animal conformations~\cite{Rubinstein1986,Cates1986,Nechaev1987,Rosa2013}, which imply self-similar, fractal conformations and dynamics, very different from the reptation of entangled linear polymers. More recent models relaxed the assumption of strict double-folded states~\cite{Michieletto2016softmatter,Ge2016,Smrek2019} and studied the appearance of transient opening of the double-folded structure thereby forming loops~\cite{Michieletto2016softmatter}. These local openings can themselves accommodate double-folded contour of the same or other neighbouring rings; these ``threadings'' do not violate the topological invariance of the system and have been found to be abundant in entangled solutions of rings~\cite{Michieletto2014acsmacro,Lee2015,Tsalikis2016}. Threading of rings is an architecture-specific TC that is not present in systems of linear chains and gives rise to unique dynamical states; for instance, inter-chain threadings can slow down rings in the bulk~\cite{Kapnistos2008,Lee2015,Doi2015}, yield a heterogeneous, glass-like dynamics~\cite{Michieletto2016pnas,Michieletto2017prl,Smrek2020} and cause anomalous response to extensional flow~\cite{Huang2019,OConnor2019} while intra-chain threadings (self-threading) can dramatically increase the relaxation time of rings in dilute conditions~\cite{Michieletto2014RCS,Soh2019}.

While it is now clear that threading between rings plays a crucial role in the dynamics of rings, its unambiguous identification remains far from trivial. The reason for this is that two rings that are mutually threaded are topologically equivalent to non-threaded ones. At present, three methods have been proposed to detect threading in solutions of rings: (i) using a local deformation of their contours and a background mesh~\cite{Michieletto2014acsmacro}, (ii) reducing their contours to primitive paths and using geometrical analysis on these paths~\cite{Tsalikis2016} and (iii) calculation of their minimal surface and analysis of intersections~\cite{Smrek2016}. 
Albeit these methods give qualitatively similar results, they are cumbersome to implement and require arbitrary choices of parameters. Additionally, they are ring-specific and cannot be mapped onto systems of linear chains, thus preventing a direct comparison between TCs in solutions of ring and linear polymers.

To overcome this problem, in this paper we propose an unambiguous and parameter-free algorithm that employs persistence homology (PH)~\cite{hatcher2002algebraic,Hiraoka2016,Nakamura2015}. We apply this algorithm to datasets from MD simulations of entangled solutions of rings and show that it allows us to define a new and broader class of TCs between ring polymers which we call ``homological threadings'', or H-threadings. Unlike previous definitions of threadings, a H-threading is defined by the homological property of the system, and its identification is unique. Persistent homology provides us with a powerful tool to extract useful information hidden in the big data obtained from, e.g., large-scale MD simulations of polymer melts~\cite{Kremer1990,Halverson2011dynamics} or X-ray tomography of elastic bands~\cite{Gomez2020}. Additionally, it has several advantages over other methods: (i) ability to detect and quantify so-far elusive hierarchical loops in the folding of rings, (ii) very efficient and stable numerical (open-source) implementation, (iii) uniqueness of the analysis result (no free parameters) and (iv) can be readily generalised to detect higher-dimensional structures in data sets. 

Importantly, while hierarchical loops are now central in state-of-the-art models to describe the static and dynamic properties of ring polymers in dense solutions~\cite{Ge2016,Michieletto2016softmatter}, their characterisation remains very poor due to a lack of algorithms that can detect them. Here we show that the PH analysis is perfectly positioned to overcome this current limitation. Finally, while in this work we focus on entangled ring polymers, our method may be employed more broadly to characterise TCs in generic polymeric materials. 

\begin{figure*}[!htb]
    \centering
    \includegraphics[width=\textwidth]{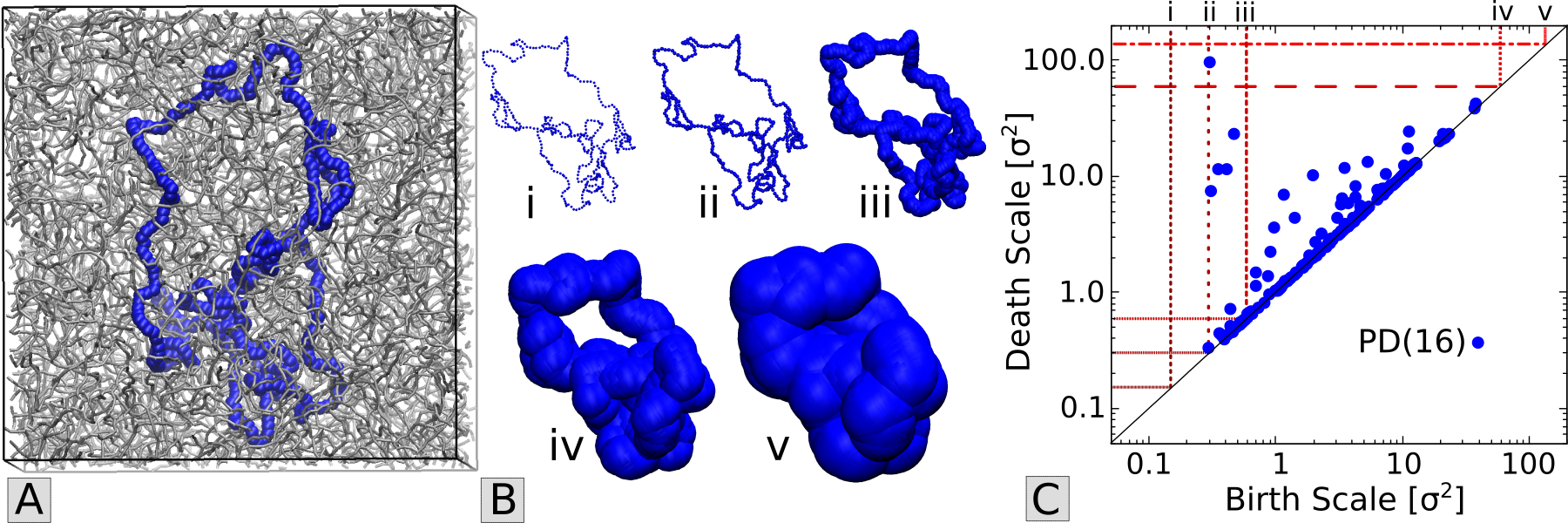}
	\vspace{-0.5cm}
	\caption{{\bf Persistent Diagram of an unknotted  and uncatenated polymer ring composed by $N=512$ monomers.} (A) Snapshot of entangled solution of unknotted and non-concatenated rings ($N=512$, $M=40$). As example, we select an arbitrary ring in this solution. (B) We consider ball representations of ring conformation with various ball radius $r(\alpha) = \sqrt{\alpha}$. (C) Persistent diagram of the ring.  Birth and death scales for five initial loops $c_k^{(m)}$ with $k=1$ to $5$ are $(b_{k}^{(m)}, d_{k}^{(m)})=(0.296, 0.331), (0.298, 96.232),(0.305, 7.393),(0.355, 11.480),(0.363, 0.440)$. Note the correspondence in numbering between (B) and (C)}.
    \label{F1}
	\vspace{-0.3cm}
\end{figure*}

\section{Materials and Methods}

\subsection*{Molecular Dynamics Simulations}
Here we analyse MD simulations reported in previous works (the interested reader may find more details in Refs.~\cite{Michieletto2016pnas,Michieletto2017prl}). Briefly, the polymers are modelled as Kremer-Grest chains made of beads of size $\sigma$. The beads are connected along the backbone by finite-extension-nonlinear-elastic (FENE) bonds and interactions between beads are governed by the Weeks-Chandler-Andersen (WCA) potential (purely repulsive Lennard-Jones). We also introduce polymer stiffness by including a Kratky-Porod term acting on triplets of neighbouring beads $U_{\rm bend} = (k_BT l_p/\sigma)(1 + \cos{\theta})$, where $\theta$ is the angle formed by consecutive bonds and $l_p=5 \sigma$.

The motion of the beads is evolved via a Langevin equation that couples the motion of the beads with an implicit solvent. The damping constant is set to $\gamma=\tau_{MD}^{-1}$ where $\tau_{MD}=\sqrt{m\sigma^2/\epsilon}$ is the microscopic characteristic time. The thermal noise satisfies the fluctuation dissipation theorem and to integrate the equations of motion we employ a velocity-Verlet algorithm with timestep $dt=0.01\tau_{MD}$.

\subsection*{Persistence Homology Algorithm} 
\label{mat_meth_PH}

Using configurations from MD simulations as an input data, the PH analysis generates a diagram called the persistent diagram (PD), from which we can extract hidden topological features \cite{hatcher2002algebraic,Hiraoka2016,Nakamura2015}.
For our present purpose, let us consider the $m$-th ring ($m \in [1,M]$); the input data is $\{ {\vec r}^{(m)}_n \} = ({\vec r}^{(m)}_1, {\vec r}^{(m)}_2, \cdots, {\vec r}^{(m)}_N)$, where ${\vec r}^{(m)}_n$ is the position of $n$-th monomer in the $m$-th ring. Associated to each monomer's coordinate we assign a ball of radius $r(\alpha) = \sqrt{\alpha}$, which should not be confused with the real radius $\sigma/2$ of the monomers. Initially, we set $\alpha=0$, so the input data is regarded a collection of volumeless points. This method could be generalized to copolymer setting a different initial radius for different types of monomers. We then gradually inflate the balls and for each value of $\alpha$ we connect all the balls that are even partially overlapping. This algorithm generates sets of topological structures made of connected balls that evolve with the balls' radius $r(\alpha)$.

\section{Discussion}

\subsection*{Persistent Homology}

Persistent homology is a method to compute the topological features of a system at different length scales~\cite{hatcher2002algebraic,Hiraoka2016,Nakamura2015}. The main concepts in this type of analysis are: the persistent homology point, that defines the size at which a certain topology of the system appears/vanishes (birth/death scale), and the persistent diagram (PD), a collection of all the PH points of fixed dimension (see Materials and Methods). Our analysis is focused on entangled solutions of unknotted, unconcatenated ring polymers; in order to capture the constraints related with the ring topology, we consider the homology of dimension 1 that identifies the formation of ``loops''. Importantly, here and in what follows, we refer to a ``loop'' as a one-dimensional (1D) geometric object detected in the PH analysis, while the term ``ring'' is reserved for the full polymer contour length (for more details see Materials and Methods). 

An example of this analysis is shown in Fig.~\ref{F1}. Starting from a single snapshot from MD simulations of ring polymers -- made as sterically interacting connected beads (see Materials and Methods) -- we associate at each bead position a sphere of size $r(\alpha)=\sqrt{\alpha}$. When $\alpha < \alpha_{min}$ ($\simeq 0.3$ in the example of Fig.~\ref{F1}Bi) there is no loop present because there are no overlapping spheres. At $\alpha = \alpha_{min}$ all the balls are connected with the consecutive along the ring and therefore the ring itself is identified as a loop in the PD analysis (Fig.~\ref{F1}Bii). It should be mentioned that the first loop to be identified by the PH algorithm is not necessarily the full ring contour length and may instead take up only a fraction of the whole ring, in agreement with previous findings using ``contact maps''~\cite{Michieletto2016softmatter,Likhtman2014b}. For increasing values of $\alpha$, certain existing loops are annihilated while others are formed (Fig.~\ref{F1}C, D) and, eventually -- at $\alpha \simeq 96$ for this particular example -- the inner space of the ring is filled in and no loop is detected. 

If a loop of connected balls exists for a given $\alpha$, a small inflation may lead to a change of topological structures yielding disappearance of such a loop and appearance of others. The sequence of changes in these 1D topological features for different $\alpha$ is intimately associated to the hierarchical, multi-scale folding of the rings, and is encoded in the PD generated by the PH algorithm. Below, we explain how to translate the PD into useful information about the folding and topological interactions between rings in solution.

For a given ($k$-th) loop on the $m$-th ring, i.e. $c_k^{(m)}$, its birth and death scales are denoted as $b_k^{(m)}$ and $d_k^{(m)}$ and capture the length-scales at which the loop first appears and then disappears, respectively. It should be highlighted that the birth scale is also related to the maximum spatial distance between two monomers that generate a certain loop; indeed, when $\sqrt{\alpha}$ is larger than the distance of two monomers these are then ``in contact'' and a loop is created by the connected portion of contour length between these two monomers. The minimum length to achieve this is $\sqrt{b_k^{(m)}}$. The death scale is instead related with the geometric 3D size of such a loop. More precisely, $\sqrt{d_k^{(m)}}$ corresponds to the radius of the maximum sphere that can be drawn inside the loop $c_k^{(m)}$.

\begin{figure*}[!tb]
\centering
	\includegraphics[width=\textwidth]{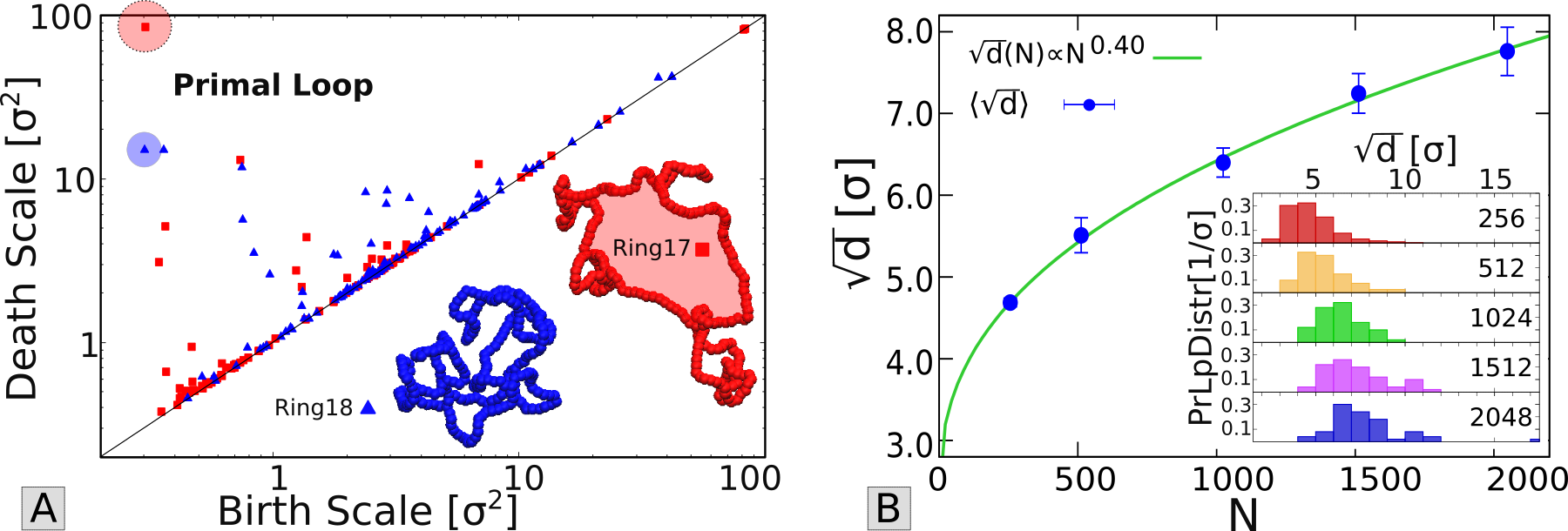}
    \vspace{-0.4cm}
	\caption{{\bf Primal Loop}. (A) Comparison of PDs obtained from different ring conformations with N=256, in this case, Ring n.~17 (red) and Ring n.~18 (blue). The largest loop (highlighted) in the red ring has a PD point with life span much larger than all the other loops.  The crumpled conformation of the blue ring has a primal loop whose life span is much shorter tha the red one.
	(B) Dependence of the primal loop size (measured as the square root of its death scale $\sqrt{d_{PL}}$) on ring length. The data for the ring length $N=256$, $512$, $1024$, $1512$, $2048$ (blue dots) are fitted with a power law curve (green curve) with exponent $\nu=0.40\pm0.11$. The inset shows the probability distribution for each ring length; notice that the peak values also show an analogous power law dependence with $N$.
	}
	\vspace{-0.4cm}
  \label{F2}
\end{figure*}
  
The PD of the $m$-th ring is made by the collection of all of the birth and death scales $(b_k^{(m)}, d_k^{(m)})$ of all the loops generated by ranging $\alpha \in [0, \infty)$, i.e. 
$PD(m) \equiv \left\{ (b_k^{(m)}, d_k^{(m)}) \in \mathbb{R}^2 | k=1,2, \dots \right\}$. 
This set of points is typically represented in the top half triangle of a two-dimensional plane as in Fig.~\ref{F1}C.

\subsection*{Persistence diagram of crumpled rings}

For each point ($b_k^{(m)}, d_k^{(m)}$) in the PD (see Materials and Methods), we define its life span $l(c_k^{(m)}) \equiv d_k^{m}- b_k^{m}$, which measures the distance from the diagonal line to the point. The points that lie far from the diagonal, i.e. with large life span, are the ones reflecting robust and persistent features of the system, whereas the ones lying close to the diagonal are often associated with noise~\cite{Kusano2017}.
Among other off-diagonal points, we call the loop associated to the point with the largest life span ``primal''. More specifically, for open non-crumpled conformations, the primal loop may corresponds to the ring itself and its birth scale is expected to be on the order of $\simeq (\sigma/2)^2$ ($\sigma$ is the size of a simulated bead). In general, the primal mirrors the presence of a large opening of the double-folded structure in the ring's conformation and its life span quantifies the size of the opening.

For example, in Fig.~\ref{F2} we show two typical conformations and their respective PDs in red (ring 17) and blue (ring 18).  The primal loop of the red ring has a large death scale and its life span is much longer than that of the blue ring (notice the logarithmic scale in the plot). This difference reflects the fact that while ring 17 displays a large opening, ring 18 assumes a more crumpled and double folded conformation entailing a much shorter primal life span (see the snapshot in Fig.~\ref{F2}A). Another notable feature of Fig.~\ref{F2}A is that ring 18 displays more points away from the diagonal than ring 17 (compare, e.g., the interval $[0.5,5] \in $ birth scale). This feature of the PD reflects the crumpled nature of the conformation: the more crumpled is a ring, the more balls of a certain radius form bridges between two distal segments of the ring contour in turn generating a moderately robust loops.

Focusing our attention on the primal loop, it is possible to study the average linear size $\sqrt{d_{PL}}$ of the loop at different ring lengths, where $d_{PL}$ is the death scale of the primal loop. In Fig.~\ref{F2}B, we show that this quantity follows a power law dependence similar to the radius of gyration. This result agrees with the existing theories~\cite{Grosberg2013,Ge2016,Sakaue2018,Sakaue2011,Halverson2014,Cates1987}, that predict an exponent $1/2 \leqslant \nu \leqslant 1/3$, with $\nu=1/2$ for the short ring length, an intermediate regime with $\nu=0.4$~\cite{Cates1987} and a compact long ring regime $\nu=1/3$~\cite{Grosberg1993}. In the inset is reported the probability distribution of $\sqrt{d_{PL}}$ for different ring size, to show that not only the average but also the peak value of the distribution is shifted towards larger sizes. For comparison, we show the probability distribution of the radius of gyration $R_g$, and its correlation with $\sqrt{d_{PL}}$ in Appendix A.

Having identified features of the PH analysis that characterise the conformation of single rings, we now aim to employ this tool to detect and quantify inter-ring interactions called threadings~\cite{Michieletto2014acsmacro} for which an unambiguous and unique (i.e. without arbitrary and free parameters) identification algorithm is needed.  

\subsection*{Topological Interactions: Homological threading}

\begin{figure*}[!hbt]
\centering
	\includegraphics[width=\textwidth]{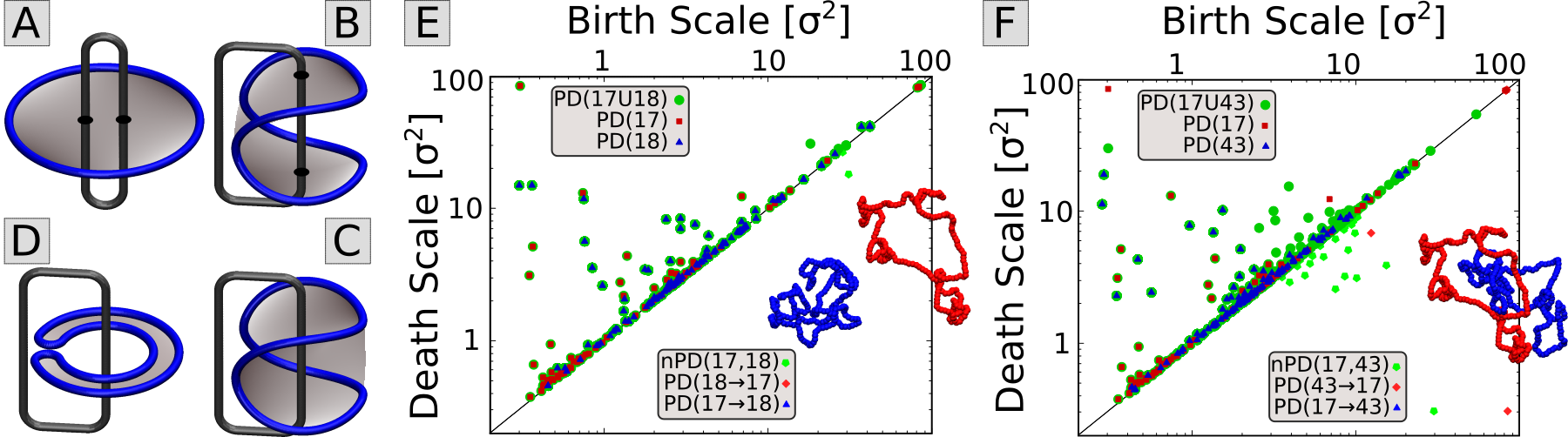}
	\vspace{-0.4cm}
	\caption{{\bf Homological threading} Schematic illustrations of homological threading (left): (A) Elongated black ring is passing twice through a planar disk; (B) One side of the black rectangular ring is passing through the curved surface spanned over the baseball seam edge; (C) Configurations of bar and the baseball seam edge is the same as those in (B), but now the curved surface is spanned in a different way; (D) A side of the black ring is passing through a gulf formed by a double-folded ring.
	PD based H-threading analysis (right): PDs of two rings ${\rm PD}(i)$ (red), ${\rm PD}(j)$ (blue), and PD obtained from the union of two rings, i.e., twin configuratio ${\rm PD}(i \cup j)$ (green). Plot E shows the PDs of two rings ($i=17$ and $j=18$) that have no H-threading between them, this is evident in the PD since all the red and blue points are overlapped by green points, i.e., there are no lost points. In the lower part of the graph are shown the new and the lost PH points but, for what said before, no lost point and only few new point at large scale are present (the two axis are exchanged). Snapshot at bottom right provides a visible inspection of the spatial position of the two rings. Plot F instead shows the ring $j= 43$ that is H-threading into $i=17$ as mirrored by the presence of lost points, i.e., the red point in top left (primal loop of ring $i$) is not overlapped by a green one (the snapshot confirm this H-threading event). The red point in the lower side correspond to this H-threading, and with them a cloud of new points are formed.}
	\label{F3_Threading}
	\vspace{-0.4cm}
\end{figure*}

Let's consider two unconcatenated and unlinked rings in 3D space; it is natural to define the arrangement of the two rings depicted in Fig.~\ref{F3_Threading}A as one in which the black ring is ``threading'' the other by crossing twice the surface spanned by the closed blue curve. Bending the blue ring into a baseball-like shape (Fig.~\ref{F3_Threading}B), the black ring still seems to pierce twice through the surface. In contrast one may argue that there is no threading in Fig.~\ref{F3_Threading}C, where the arrangements of the curves is the same but the spanned surface is forming a gulf that is not pierced by the former curve. Given the symmetry of the baseball seam curve, which side is selected as a spanning surface is a subtle matter. Such a subtlety would be even more enhanced for 3D crumpled rings. Additionally, while the situations in Figs.~\ref{F3_Threading}B and D appear different, they can be continuously mapped to the same three dimensional arrangement (as Fig.~\ref{F3_Threading}D); furthermore, the physical hindering of the black curve on the dynamics of the blue may be similar. Finally, in both cases, the obstacle can be removed by a continuous deformation, for instance by sliding it to left direction, as threadings are not conserved under isotopy.
Note that the figures Figs.~\ref{F3_Threading}A-D are a typical example of isotropic change, and there can be other examples with similar ambiguity when we define threading in an intuitive way.

Nevertheless, here we aim to propose a definition for which all arrangements (Fig.~\ref{F3_Threading}A-E) are identified as threadings and, because discovered through a persistence homology analysis, we will dub them ``homological threadings'' (H-threadings). 

In order to characterize the topological interaction between two rings, say $i$ and $j$, we construct four PDs from the configuration data of the two rings. Two of them are the PDs of each ring, which we write PD($i$) and PD($j$). The third one is the PD obtained from the union of the configurations of these two rings PD($i \cup j$) and the last one is the union of PD($i$) and PD($j$) which we denote as uPD($i,j$) $\equiv$ PD($i$) $\cup$ PD($j$). From these, we create the following PDs as set difference
\begin{eqnarray*}
&&{\rm PD}(j \rightarrow i) \equiv {\rm PD}(i) \setminus {\rm PD}(i \cup j)\ , \\
&&{\rm nPD}(i,j) \equiv  {\rm PD}(i \cup j) \setminus {\rm uPD}(i,j) \ ,
\end{eqnarray*}
where ${\rm PD}(j \rightarrow i)$ represents PD points which are associated to a loop in ring $i$ that is however {\it lost} in the PD in which both configurations are included. Such a loop, identified as a lost point of ring $i$, should be interpreted as a loop being threaded by ring $j$. Indeed, if any segment of ring $j$ is passing through an area that was identified as an opening in ring $i$'s conformation, then it must represent an obstacle for the ring $i$ in the sense described in Fig.~\ref{F3_Threading}. By symmetry, the same applies to ${\rm PD}(i \rightarrow j)$ with the interchange between $i$ and $j$. These relations constitute our definition of {\it H-threading} events. At the same time, ${\rm nPD}(i,j)$ represents {\it new} points created by considering both conformations and that are absent in the union of the PDs obtained using isolated conformations; these points represent new loops that are formed when balls of a certain radius bridge segments belonging to different rings and therefore may also be considered as obstacles, restricting the conformational freedom of the individual rings $i$ and $j$.

\begin{figure*}[!t]
\centering
	\includegraphics[width=\textwidth]{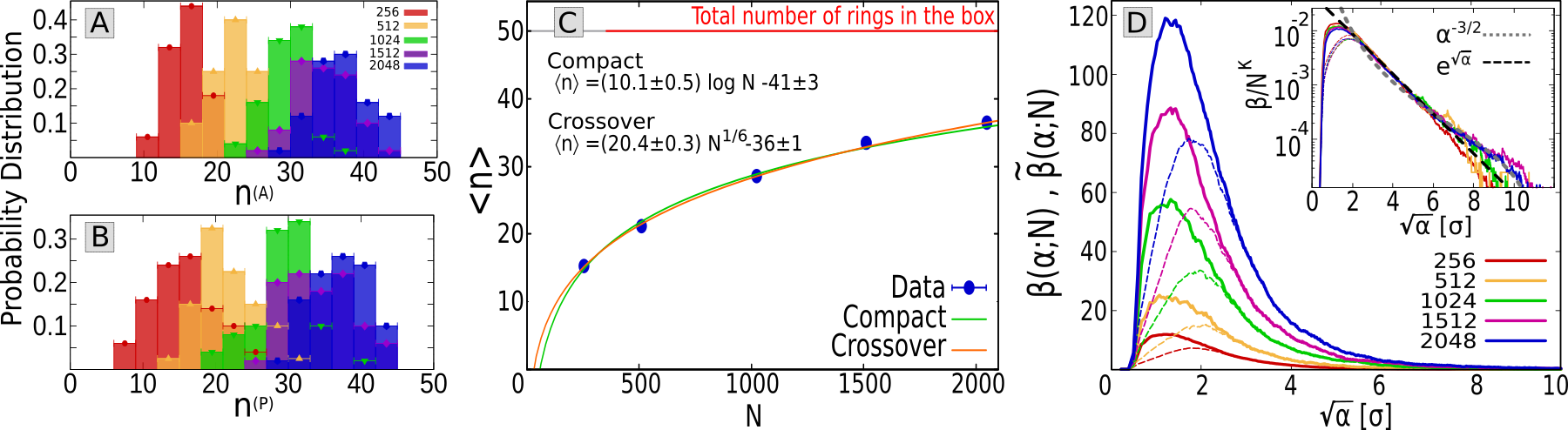}
	\vspace{-0.4cm}
	\caption{{\bf Threading statistics.} Probability distributions of active (A) $n^{(A)}$ and passive (B) $n^{(P)}$ threadings for various ring length $N$. In (C), the average of these distribution $\langle n \rangle = \langle n^{(A)} \rangle =  \langle n^{(P)} \rangle $ is shown as a function of ring length. The data are fitted with $\langle n \rangle \propto N^{1/6}$ (crossover $\nu \simeq 0.4$) and $\langle n \rangle \propto \log N$ (compact $\nu = 1/3$). Notice that these values are computed as number of H-threadings per single ring meaning that every one of our longest rings are H-threading more than half other rings in the box.
	(D) The Betti Number is another measure to obtain information on TCs. In the main plot we show the Betti number for isolated rings (solid curves) $\beta(\alpha ; N)$ as calculated from the PD of ring conformations for various ring length $N$ and we compare it with the quantity ${\tilde \beta}(\alpha ; N)$ calculated from Eq.~(\ref{betti_threaded_loops}) (dashed lines, see text for details). The inset shows a log-linear plot of the same curves rescaled by the factor $(N/N_0)^\kappa$. After the peak, all the data are well described by the fit $ \beta (\alpha ; N) = {(\frac{N}{N_0})}^\kappa \exp{ ( -\sqrt{\frac{\alpha}{\alpha_0}} )}$ (black dashed line) with $\kappa=1.2\pm0.1$, $N_0=12\pm3$ and $\alpha_0=1.4\pm0.2$. The collapse of these curves suggest that (H-threaded) loops display a super-linear dependence on $N$ and an exponential decay with $\alpha$. Also shown is a power-law function $f(\alpha) \sim \sqrt{\alpha}^{-3}$(gray dotted curve); see a discussion in Appendix B for the loop size statistics. }
  \label{F_threading_dist}
  \vspace{-0.4cm}
\end{figure*}

Our definition of H-threading respects the directionality of this interaction: points in ${\rm PD}(i \rightarrow j)$ identify arrangements in which ring $i$ is actively H-threading $j$ while points in ${\rm PD}(j \rightarrow i)$ find ones in which $i$ is passively H-threaded by $j$. To quantify these events, we can count the number $n^{(A)}$ of rings through which a reference ring is threading (active threading) and the number $n^{(P)}$ of ring threading into a reference ring (passive threading); their statistics are shown in Fig.~\ref{F_threading_dist}A-B. Due to symmetry, $\langle n \rangle = \langle n^{(A)} \rangle =  \langle n^{(P)} \rangle $, and we find that it grows with ring length as $\langle n \rangle\propto N^{1/6}$ (or $\langle n \rangle = \ln N$ see Fig.~\ref{F_threading_dist}C and discussion below).
To motivate such a scaling law, let us count the number of surrounding rings, which have contacts with the reference ring. This quantity, i.e., the coordination number, is estimated as $\sim R^3 M/V = \rho R^3/N \sim N^\alpha$ with $\alpha=3\nu-1$, where $R\sim N^\nu$ is the ring spatial size. The use of effective exponents $\nu=2/5$, valid in intermediate cross-over $N$ regime, we find $\alpha=1/6$. On the other hand, the compact scaling exponents $\nu=1/3$ is expected for asymptotically long rings, then, we get $\alpha=0$, implying a logarithmic dependence on $N$. Smrek and Grosberg observed a similar chain length dependence  for the number of rings penetrated by a single ring in their minimal surface analysis~\cite{Smrek2016}.
The distinction between $n^{(A)}$ and $n^{(P)}$ shows up in their distributions, with larger width for $n^{(P)}$. This may reflect the fact that passive threadings are more sensitive to the opening of loops, i.e., larger $n^{(P)}$ for rings having a primal loop with longer life span, and vice versa.

We finally stress that since $\langle n \rangle$ is the average number of H-threadings \textit{per ring}, our findings strongly suggest that our systems of rings are percolating, in the sense that all rings are connected to each other through H-threadings.

\subsection*{Betti number}
So far, we have analyzed the H-threading statistics per ring as a function of ring length. However, it is natural to explore the statistics of H-threadings as a function of the loop size into which the H-threading takes place. To this end, we define the Betti number 
\begin{eqnarray*}
\beta^{(m)}(\alpha ; N) \equiv \int_{\alpha}^{\infty}{\rm d}d \int_0^{\alpha}{\rm d}b \sum_{k} \delta(b-b_k^{(m)}) \delta(d-d_k^{(m)}) \label{eq:betti}
\end{eqnarray*}
 for a given configuration of the $m$-th ring $\{ {\vec r}_n^{(m)}\}$ with monomer index $n \in [1,N]$.

The Betti number $\beta^{(m)}(\alpha; N)$ counts the number of loops in the $m$-th ring if observed at the spatial resolution $\sqrt{\alpha}$. In Fig.~\ref{F_threading_dist}D, we compare the average Betti number per ring
\begin{eqnarray}
\beta (\alpha ; N) = M^{-1}\sum_{m=1}^M \beta^{(m)}(\alpha ; N) \, , \label{ave_betti}
\end{eqnarray}
with the quantity
\begin{eqnarray}
{\tilde \beta}(\alpha ; N) \equiv M^{-1}\int_{\alpha}^{\infty}{\rm d}d \int_0^{\alpha}{\rm d}b \sum_{k} \delta(b-{\tilde b}_k) \delta(d-{\tilde d}_k) \, ,  \label{betti_threaded_loops}
\end{eqnarray}
where $({\tilde b}_k, {\tilde d}_k)$ are the birth and death scales of the loops associated with all H-threading events, i.e.,
\begin{eqnarray}
({\tilde b}_k, {\tilde d}_k)\in \bigcup_{i \neq j} 
\left\{(b_l,d_l) \in PD(j \rightarrow i) \right\}  \label{threaded_loops} 
\end{eqnarray}

Therefore, while Eq.~\ref{ave_betti} represents the total number of loops per ring (averaged over rings) if observed at the spatial resolution $\sqrt{\alpha}$, Eq.~\ref{betti_threaded_loops} counts only loops which are associated with a H-threading event. 

Mathematically, the set the element of which is $(\tilde b_k,\tilde d_k)$ is no longer a PD in any sense because the right-hand side of Eq. \ref{threaded_loops} is defined not as a disjoint union but a union.
Only for our analysis, the difference between union and the disjoint union does not affect the original PD analysis because all multiplicity value in $PD(i)$ is confirmed to be unity numerically.

By comparing these two quantities (see Fig.~\ref{F_threading_dist}D, solid and dashed lines, respectively) one can notice that:  (i) The number of threaded loops approach the total number of loops from below, and these two quantities converge at large enough $\alpha$; (ii) the scaling function $f(\alpha) \sim e^{- \sqrt{\alpha/\alpha_0}}$ well captures the scale-dependence of the Betti number except within an uninteresting small $\alpha \sim$ monomer scale. There is a small deviation from the exponential behavior at $\sqrt{\alpha} \gtrsim 7$. This may be a signal to a self-similar loop statistics as conjectured in loopy globule model~\cite{Ge2016} (see also Appendix) ;
(iii) the Betti number follows a super-linear scaling with ring length $\beta (\alpha ; N) = N^{\kappa} f(\alpha)$ with exponent $\kappa \simeq 1.2 \pm 0.1$, indicating that the number of (threaded) loops grow at least extensively with the rings' contour length (see Fig.\ref{F_threading_dist}D inset). 

We note that this finding is compelling evidence for the argument that even if the number of neighbouring rings H-threaded at any one time is bounded to plateau in the asymptotic limit (Fig.~\ref{F_threading_dist}C) the number of H-threaded loops (and hence of the induced TCs) increases at least linearly with the rings' contour length. 

\section*{Conclusions and perspectives}

The notion of ``loop'' is often invoked as a structural motif in solutions of ring polymers~\cite{Smrek2015a,Smrek2019a,Ge2016}, but its precise definition is far from evident. We have proposed an operational definition of ``loop'' and associated topological constraints, here dubbed H-threadings, based on a persistent homology (PH) algorithm, which allows us to unambiguously quantify the statistics of loops in entangled solution of nonconcatenated rings. We have discovered that our PH analysis naturally yields (i) a sublinear scaling $\sqrt{d_{PL}} \sim N^{0.4}$ of the primal loop's death scale, reflecting the known scaling statistics of rings' size $R_g \sim N^{2/5}$ in the crossover regime~\cite{Cates1987} (Fig.~\ref{F2}), (ii) a slow increase of H-threaded neighbours (Fig.~\ref{F_threading_dist}C) and (iii) a superlinear scaling of the Betti number -- $\beta (\alpha ; N) = N^{\kappa} f(\alpha)$ with $\kappa \simeq 1.12$ -- which quantifies the number of loops at length-scale $\sqrt{\alpha}$ (Fig.~\ref{F_threading_dist}D, see also Appendix B).

It is important to stress that quantifying the statistics of loops in folded rings is not trivial~\cite{Rosa2013,Michieletto2016softmatter,Smrek2019a}; recent theories rely on the notion that loops are formed on multiple scales in crumpled rings~\cite{Ge2016} but lack a precise way to quantify their abundance and length-scale dependence. Here we provide both via a physically appealing and unambiguous method (Fig.~\ref{F_threading_dist}D). 

The key point of the present work is that the number of rings that are H-threaded by any one ring increases slowly (as $N^{1/6}$ in the crossover regime and as $\log{N}$ in the asymptotic limit) with contour length (Fig.~\ref{F_threading_dist}C). This is likely due to the fact that the number of neighbours is itself conjectured to saturate around the so-called Kavassalis-Noolandi number, i.e. 10-20~\cite{Rosa2013,Ge2016,Sakaue2018,Sakaue2011}. Indeed, we note that $\langle n \rangle$ follows a similar molecular weight dependence to that of the effective ``coordination number'' that has been studied in numerical simulations~\cite{Rosa2013} and plays an important role in phenomenological mean-field theories~\cite{Sakaue2018,Sakaue2011,Sakaue2019,Sakaue2016}.
On the other hand, by analysing the behaviour of the Betti number we discovered that the number of (H-threaded) loops increases at least linearly with the rings' length, in turn implying an extensive growth of threading TCs also in the asymptotic limit (Fig.~\ref{F_threading_dist}D, inset). We argue that this observation is related to the fact that crumpled ring polymers, unlike the ordinary compact globules, display a very ``rough'' surface whereby an extensive number of the segments are exposed and prone to be in contact with the neighbouring rings~\cite{Grosberg2013}.  

In closing, we list some perspectives for the proposed method. The most natural is to investigate the relation between the present results and the rheological properties of the system. One way to proceed in this direction is to trace the time evolution of PDs. For instance, the time evolution of lost and new points in PDs would allow us to analyse the \textit{temporal} persistence of TCs between pair of rings and relate it to the size of the involved H-threadings and loops. Second, our definitions of loop and H-threading can be applied to linear polymer system too. It would be interesting to analyse entangled linear polymer solutions, and see how the results compare to the present ones for ring polymers. In fact, the portability of our method to systems of polymers with other topologies is an advantage of our approach over others that are only tailored for ring polymers. Third, this method can be readily generalised to solutions of co-polymers by setting a different initial radius for different types of monomers. 

Finally, we would like to emphasize that the proposed method of analysis relies on the construction of new PDs, such as ${\rm PD}(j \rightarrow i)$ and ${\rm nPD}(i,j)$ from a pair of PDs. This is not a standard procedure in PH analysis and it is introduced here for the first time. We expect that this idea has a significant potential in the broader application of persistent homology analysis not only in polymer systems but also in other problems, e.g. to study the glass transition~\cite{Nakamura2015}.

\begin{acknowledgments}
TS is supported by JSPS KAKENHI (No. JP18H05529) from MEXT, Japan, and JST, PRESTO (JPMJPR16N5). TN is supported by JSPS KAK-ENHI Grant Number 15K13530 and JST-PRESTO Grant Number JPMJPR15ND, Japan.
DM is supported by the Leverhulme Trust through an Early Career Fellowship (ECF-2019-088). 
The \href{https://www.wpi-aimr.tohoku.ac.jp/hiraoka_labo/homcloud/index.en.html\#download}{\textit{\underline{HomCloud}}} software used for the PH analysis is freely available and distributed under GPL3.
\end{acknowledgments}

\section{Appendix A: Comparison of primal loop size and radius of gyration }

\begin{figure*}[htp]
\centering
\includegraphics[width=0.9\textwidth]{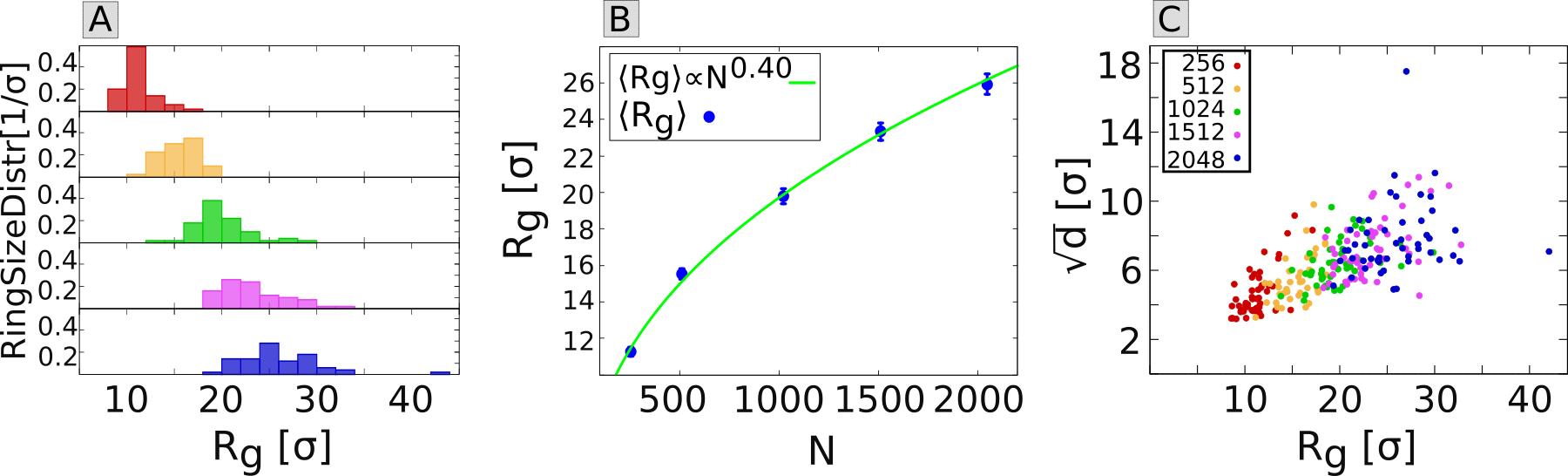}
\caption{{\bf Ring size.} (A) Distribution of the radius of gyration ($R_g$) for various ring lengths $N$ (red: $N=256$, yellow:$N=512$, green:$N=1024$, purple:$N=1512$ and blue:$N=2048$). (B) The average ring size scale with the same power law $\langle R_g \rangle \sim N^{0.4}$ as in the case of the Primal Loop size (see Fig.~\ref{F2}).
(C) Primal loop size of each ring is plotted against the radius of gyration (same color code as A). The plot shows that while these two quantities are correlated on average, they represent distinct aspects of the ring shape. }
\label{Appendix_Fig1}
\end{figure*}

In Fig.~\ref{Appendix_Fig1}, we show the probability distribution of the gyration radius $R_g$ for various ring lengths. The overall trends, including the shape of distribution and its chain length dependence, are similar to those for the primal loop size (Fig.~\ref{F2}B). This does not mean, however, the large ring always has the large primal loop, as shown in the correlation plot in Fig.~\ref{Appendix_Fig1}C.

\section{Appendix B: Loop size distribution}

\begin{figure*}[htpb]
\centering
\includegraphics[width=0.9\textwidth]{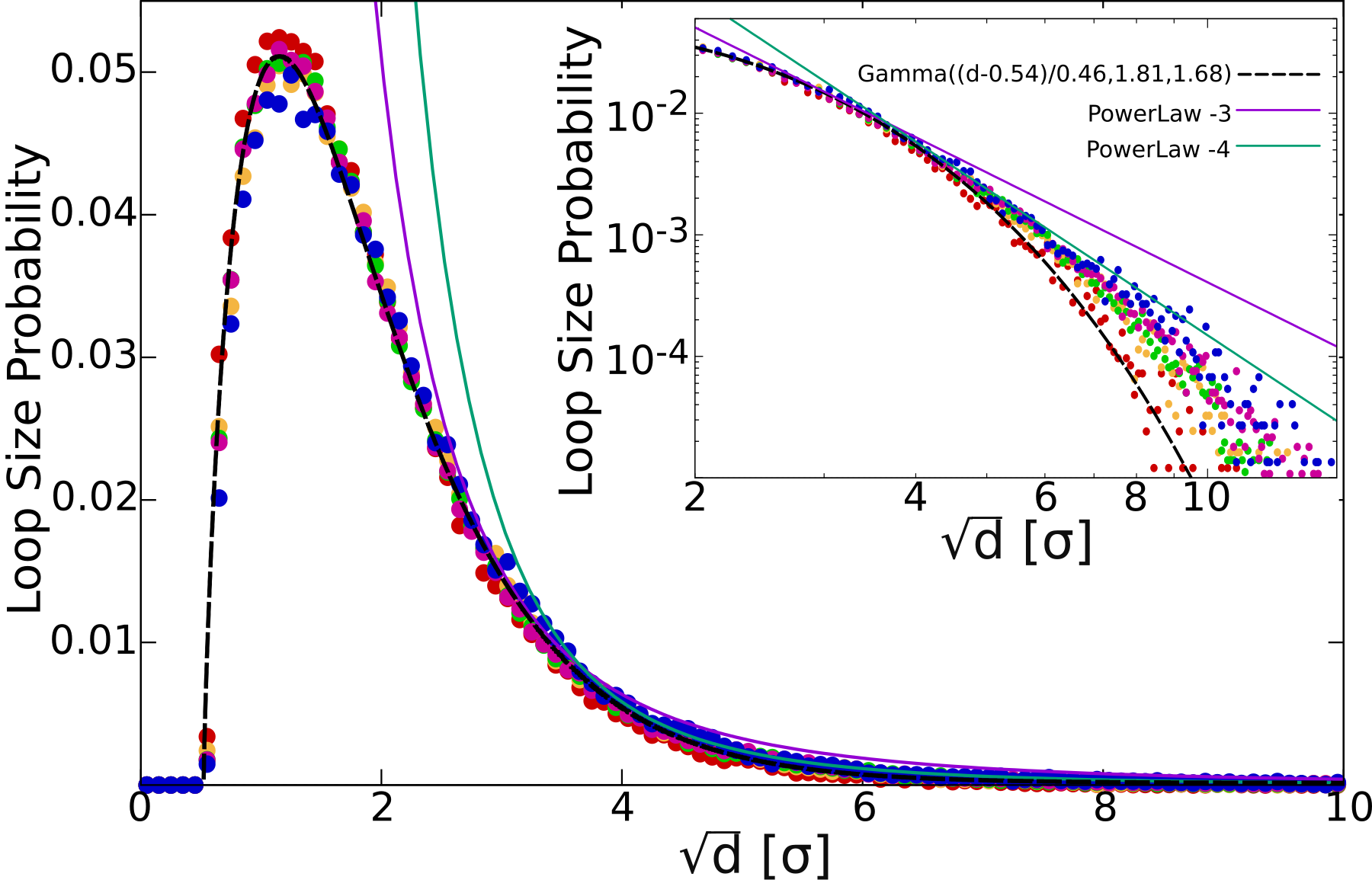}
\caption{{\bf Loop size distribution} ${\tilde n}_d(r)$ at different ring lengths  (red: $N=256$, yellow:$N=512$, green:$N=1024$, purple:$N=1512$ and blue:$N=2048$). Using the square root of the death scale as measure of the size of a single loop, we observe that the probability to form a loop of size $\sqrt{d}$ follow the well-known gamma distribution with a shift $p(\sqrt{d})=Gamma((\sqrt{d}^{-3}-0.54)/0.46, 1.81, 1.68)$, where $0.5$ is half the average distance between single monomers. The fit captures the peak shape and the subsequent exponential decay for all the ring size. In the tails, the fit seems to correctly describe the short ring behaviour while the longer rings experience a deviation that tend to the power law behavior $\sim r^{-4} $. The cyan and purple line represent a power law fit with exponents $\sqrt{d}^{-4}$ and $\sqrt{d}^{-3}$ respectively.}
\label{Appendix_Fig2}
\end{figure*}

In the main text, we have analyzed the loop size distribution by means of the Betti number (Fig.~\ref{F_threading_dist} D).
\begin{eqnarray}
n_{\beta}(r) \equiv  \beta(\alpha = r^2)
\end{eqnarray}
and suggested that the large scale part may be described by a power law decay $n_{\beta} (r) \sim r^{-3}$.
Another way to quantify the loop size distribution is to count the number of loops based on their death scale $r=\sqrt{d}$. This is because when loops with 3D size $r$ are filled up by the balls of size $\sqrt{d}$ they disappear (or die). This quantity can be readily computed from the persistence diagrams as
\begin{eqnarray}
{\tilde n}_d(r) \equiv  \sum_{m=1}^M \sum_k \delta(r- \sqrt{d_k^{(m)}}) 
\end{eqnarray}
As shown in Fig.~\ref{Appendix_Fig2}, while the short-scale part of ${\tilde n}_d(r)$ is well-fitted by a Gamma distribution -- which indicates the random distribution of the loop size -- at longer scales, the apparent deviation from the exponential decay shows up, and the tail seems to be better fitted by a power-law with exponent $\sim -4$. This leads to a cumulative distribution
\begin{eqnarray}
n_d(r) \equiv \int_r^{Rg} \tilde{n}_d(r) dr \sim r^{-3} \, .
\end{eqnarray}
Therefore, although more statistics and larger systems are necessary, our tentative conclusion is that both quantities $n_{\beta}(r)$ and $n_d(r)$ point to a self-similar loop size distribution conjectured in loopy-globule model~\cite{Ge2016}, i.e. decaying with the 3D size of loop as $r^{-3}$. This should not be confused with the better known decay of the number of loops as a function of the 1D separation as $\sim l^{-\gamma}$, with $\gamma \simeq 1.1$ for ring polymers~\cite{Grosberg2013}.
We again emphasize that although the term ``loop'' is often invoked in systems of ring polymers as crucial statistical structural element, its precise definition is far from trivial (hence, usually not made). Through PD analysis, we have here proposed two definitions for the loop size distribution. To clarify which one better suits for a particular purpose requires further investigations.

\bibliography{library}

\end{document}